\documentstyle[12pt,epsfig,graphics,cite,amssymb,amsmath,bm]{article}
\begin{document}\bibliographystyle{plain}\begin{titlepage}
\renewcommand{\thefootnote}{\fnsymbol{footnote}}\hfill
\begin{tabular}{l}HEPHY-PUB 921/12\\UWThPh-2012-34\\November
2012\end{tabular}\\[2cm]\Large\begin{center}{\bf EXACT SOLUTIONS
OF BETHE--SALPETER EQUATIONS WITH INSTANTANEOUS
INTERACTIONS}\\[1cm]\large{\bf Wolfgang
LUCHA\footnote[1]{\normalsize\ {\em E-mail address\/}:
wolfgang.lucha@oeaw.ac.at}}\\[.3cm]\normalsize Institute for High
Energy Physics,\\Austrian Academy of Sciences,\\Nikolsdorfergasse
18, A-1050 Vienna, Austria\\[1cm]\large{\bf Franz
F.~SCH\"OBERL\footnote[2]{\normalsize\ {\em E-mail address\/}:
franz.schoeberl@univie.ac.at}}\\[.3cm]\normalsize Faculty of
Physics, University of Vienna,\\Boltzmanngasse 5, A-1090 Vienna,
Austria\\[3cm]{\normalsize\bf Abstract}\end{center}\normalsize

\noindent The Bethe--Salpeter approach allows for
quantum-field-theoretic descriptions of relativistic bound states;
its inherent complexity, however, usually prevents to find the
exact solutions. Under suitable simplifying assumptions about the
systems discussed, we derive {\em analytically\/} examples of
rigorous solutions to the instantaneous homogeneous
Bethe--Salpeter equation by relating tentative solutions to the
interactions responsible for formation of bound states.

\vspace{3ex}

\noindent{\em PACS numbers\/}: 11.10.St, 03.65.Ge, 03.65.Pm
\renewcommand{\thefootnote}{\arabic{footnote}}\end{titlepage}

\section{Introduction}One of the great challenges in theoretical
elementary particle physics still is the description of bound
states in a way fully consistent with all requirements imposed by
special relativity and quantum mechanics, that is, within the
framework of relativistic quantum field theory. In principle, the
appropriate tool to achieve this goal is the {\em Bethe--Salpeter
formalism\/} \cite{BSE_APS,GML,BSE}: for bound-state constituents
with fixed features, the {\em homogeneous Bethe--Salpeter
equation\/} governs all the bound states. However, for various
practical reasons some of which we recall below, in its
applications to both quantum electrodynamics and quantum
chromodynamics frequently some simplified equations, situated
along a path of nonrelativistic reduction, are used. Its
cornerstones are the Salpeter equation \cite{SE} and the reduced
Salpeter equation \cite{RSE,Lucha92C}.

The nature of such bound-state equations renders difficult to
find, for given interactions of the bound-state constituents,
their exact analytic solutions. However, a knowledge of, at least,
{\em some\/} exact solutions facilitates to judge the significance
of the outcomes of reduction steps or numerical solution
techniques. In view of this, we remember a straightforward idea
used in {\em quantum physics}, e.g., by Neumann and Wigner
\cite{vNW} for studies of the following~kind.

Within the formalism of {\em nonrelativistic quantum mechanics},
bound states are described by the time-independent Schr\"odinger
equation, the eigenvalue equation of the Hamiltonian operator
controlling the dynamics of the system under consideration for
energy eigenvalues $E$ and associated eigenfunctions $\psi.$ The
{\em bound states\/} correspond to the {\em discrete\/}
eigenvalues in the spectrum of this operator. For a particle of
mass $m$ experiencing interactions induced by some potential $V,$
this Schr\"odinger equation in configuration-space representation
reads$$\left[-\frac{\Delta}{2\,m}+V(\bm{x})\right]\psi(\bm{x})=E\,\psi(\bm{x})\
,\qquad\Delta\equiv\bm{\nabla}\cdot\bm{\nabla}\ .$$Typically, the
interaction potential $V$ is inferred from physical considerations
or principles, and, for this potential, the solutions
$\{(E,\psi)\}$ of the Schr\"odinger equation then are derived.
However, for the purpose of just constructing examples for exact
solutions $(E,\psi)$ related to potentials $V$ entering into the
Schr\"odinger equation one may also follow the opposite~route: on
an equal footing, one may postulate, for a chosen value of $E,$
any preferred or convenient shape of the solution $\psi$ and see
whether one is able to find analytically the related potential
$V.$ This is easily done for vanishing binding energy $E,$ i.e.,
for $E=0,$ with the general~result
$$V(\bm{x})=\frac{\Delta\psi(\bm{x})}{2\,m\,\psi(\bm{x})}\
.$$Assuming the potential $V$ to be {\em spherically symmetric},
$V(\bm{x})=V(r),$ $r\equiv|\bm{x}|,$ the choice \cite{vNW}
$$\psi(\bm{x})=\psi(r)=\frac{\sin(a^3\,r^3)}{r^2}\ ,\qquad
a\equiv2\,m\ ,$$for a likewise spherically symmetric stationary
solution thus fixes \cite{vNW} the central
potential:\footnote{Here, the eigenvalue $E=0$ is located in the
continuous spectrum of the Hamiltonian, thus
not~discrete.}$$V(r)=\frac{2}{a\,r^2}-9\,a^5\,r^4\ .$$

We intend to adapt the above procedure, in quantum mechanics
applied to Schr\"odinger problems, to the {\em Bethe--Salpeter
approach\/}. To this end, we first identify, in
Sec.~\ref{Sec:RIBSE}, examples of Bethe--Salpeter equations with
internal structure sufficiently simple to allow for the kind of
{\em inversion\/} we have in mind. For these tractable bound-state
equations, we then, in Sec.~\ref{Sec:CSP}, relate postulated
candidate solutions to confining and non-confining
interaction~potentials.

\section{Bethe--Salpeter Formalism in Instantaneous Limit}
\label{Sec:RIBSE}Motivated by the needs of physics, the aim of the
present discussion is to construct~analytic solutions to equations
of motion describing relativistic bound states which we assume to
be composed of some fermion, with momentum $p_1,$ and some
antifermion, with momentum $p_2;$ the total momentum of the bound
state is $P\equiv p_1+p_2$ and its mass $M$ is given~by~$M^2=P^2.$

\subsection{Homogeneous Bethe--Salpeter Equation
\cite{BSE_APS,GML,BSE}}\label{Sec:BSE}In principle, within the
framework of relativistic quantum field theory the adequate tool
for the (Poincar\'e-covariant) description of bound states is the
Bethe--Salpeter formalism \cite{BSE_APS,GML,BSE}. In this approach
a bound state ${\rm B}(P)$ of momentum $P$ and mass
$M\equiv\sqrt{P^2}$ is described by a Bethe--Salpeter amplitude
defined in configuration-space representation as matrix element of
the time-ordered product of the field operators of the bound-state
constituents evaluated between vacuum state $|0\rangle$ and bound
state $|{\rm B}(P)\rangle.$ In momentum-space representation, the
Bethe--Salpeter amplitude, after splitting off the
center-of-momentum motion of the bound state and suppressing all
indices generically denoted by $\Phi(p,P),$ encodes the
distribution of the relative momentum $p$ of the two bound-state
constituents; it satisfies the formally exact homogeneous
Bethe--Salpeter equation, which involves two kinds of dynamical
ingredients: the full propagators $S_i(p_i)$ of the constituents
with individual momenta $p_i,$ $i=1,2,$ and the Bethe--Salpeter
interaction kernel $K(p,q,P),$ by construction a fully amputated
four-point Green function of the bound-state constituents defined
perturbatively by summation of the countable infinity of all
Bethe--Salpeter-irreducible Feynman diagrams for two-particle into
two-particle scattering. Skipping all indices, this
Bethe--Salpeter equation~generically~reads
\begin{equation}\Phi(p,P)=\frac{\rm i}{(2\pi)^4}\,S_1(p_1)\int{\rm
d}^4q\,K(p,q,P)\,\Phi(q,P)\,S_2(-p_2)\ .\label{Eq:BSE}
\end{equation}Unfortunately, attempts to apply the Bethe--Salpeter
formalism to relativistic bound-state problems are doomed to face
various grave obstacles, such as the impossibility to determine
the Bethe--Salpeter interaction kernel beyond the tight limits of
perturbation theory, or the appearance of excitations in the
relative time variable of the bound-state constituents, that is,
of solutions called abnormal, difficult to interpret in the
framework~of~quantum~physics.\footnote{There are, however, very
sound arguments \cite{JS96} claiming that the presence of
time-like excitations~might be just an artifact of the {\em ladder
approximation\/} to the Bethe--Salpeter kernel, incorporating the
interactions responsible for the bound states only by an iteration
of a single-particle exchange between the constituents.} They may
be remedied in {\em three-dimensional reductions\/} of the
Bethe--Salpeter equation; the most well-known of the resulting
bound-state equations is the one proposed by Salpeter \cite{SE}.

\subsection{(Full) Salpeter Equation \cite{SE}}\label{Sec:SE}In
order to facilitate the formulation of instantaneous
Bethe--Salpeter equations, let's recall the standard definition of
one-particle energy $E_i(\bm{p}),$ one-particle Dirac Hamiltonian
$H_i(\bm{p}),$ and energy projection operators
$\Lambda_i^\pm(\bm{p})$ for positive and negative energy of
particle~$i=1,2$:$$E_i(\bm{p})\equiv\sqrt{\bm{p}^2+m_i^2}\ ,\qquad
H_i(\bm{p})\equiv\gamma_0\,(\bm{\gamma}\cdot\bm{p}+m_i)\
,\qquad\Lambda_i^\pm(\bm{p})\equiv\frac{E_i(\bm{p})\pm
H_i(\bm{p})}{2\,E_i(\bm{p})}\ .$$The projection operators
$\Lambda_i^\pm(\bm{p})$ satisfy, in terms of the
charge-conjugation Dirac matrix~$C,$
$$\left[\Lambda_i^\pm(\bm{p})\right]^{\rm c}\equiv
\left[C^{-1}\,\Lambda_i^\pm(\bm{p})\,C\right]^{\rm
T}=\Lambda_i^\mp(\bm{p})\ ,\qquad i=1,2\ .$$

The reduction of the Bethe--Salpeter equation to the Salpeter
equation relies on exactly two simplifying assumptions (brief
reviews of such reductions can be found in
Refs.~\cite{Lucha91,Lucha:Oberwoelz,Lucha:Dubrovnik}):

First, the instantaneous approximation assumes that in the
center-of-momentum frame of the bound states, fixed by
$P=(M,\bm{0}),$ all interactions between bound-state constituents
are instantaneous. In this (``static'') limit, the Bethe--Salpeter
interaction kernel $K(p,q,P)$ depends only on the spatial
components, $\bm{p}$ and $\bm{q},$ of the relative momenta $p$ and
$q$ involved, that is, takes the form $K(p,q,P)=K(\bm{p},\bm{q}).$
If, furthermore, the propagators $S_i(p_i)$ of both constituents
may be assumed to be entirely free from any nontrivial dependence
on the zero component $p_0$ of the relative momentum $p,$
integrating w.r.t.\ $p_0$ reduces the Bethe--Salpeter equation
(\ref{Eq:BSE}) to a kind of instantaneous Bethe--Salpeter equation
for the~Salpeter~amplitude$$\phi(\mbox{\boldmath{$p$}})\equiv
\frac{1}{2\pi}\int{\rm d}p_0\,\Phi(p)\ .$$For various attempts in
this direction, see, for instance,
Refs.~\cite{Lucha05:IBSEWEP,Lucha06:GIBSEEQP-C7} and references
therein.

Second, any bound-state constituent is assumed to propagate as
free particle with some effective mass $m_i$ required to encompass
appropriately all dynamical self-energy effects.\footnote{This
forms the example par excellence of the trivial $p_0$ dependence
of the propagators requested~above.} In other words, any fermion
propagator $S_i(p)$ is approximated by its corresponding
free~form\footnote{The assumption of free propagators for the
bound-state constituents may, however, encounter a~serious
conceptual problem for the following reason: in quantum field
theory the infinite tower of Dyson--Schwinger equations relates
any $n$-point Green function to at least one $(m\gneqq n)$-point
Green function; this entails~that propagators, being two-point
Green functions, and the four-point Green function entering as an
interaction kernel cannot be viewed as independent. In particular,
in quantum chromodynamics (the theory describing the strong
interactions) free propagators for any of its fundamental quark
and gluon degrees of freedom are incompatible with the feature of
colour confinement, exhibited by this unbroken non-Abelian
gauge~theory.}$$S_{i,0}(p,m_i)=\frac{{\rm i}}{\not\!p-m_i+{\rm i}
\,\varepsilon}\equiv{\rm i}\,\frac{\not\!p+m_i}{p^2-m_i^2+{\rm i}
\,\varepsilon}\ ,\qquad\varepsilon\downarrow0\ ,\qquad i=1,2\ .$$

Imposing both restrictions above on the Bethe--Salpeter equation
(\ref{Eq:BSE}) yields, by contour integration in the complex-$p_0$
plane and the residue theorem, the Salpeter equation (where,
trivially, $\bm{p}\equiv\bm{p}_1=-\bm{p}_2$ and $P_0=M$ in the
center-of-momentum frame of the bound state)
\begin{align}\phi(\bm{p})&=\int\frac{{\rm d}^3q}{(2\pi)^3}
\left(\frac{\Lambda_1^+(\bm{p}_1)\,\gamma_0\,
[K(\bm{p},\bm{q})\,\phi(\bm{q})]\,\gamma_0\,\Lambda_2^-(\bm{p}_2)}
{P_0-E_1(\bm{p}_1)-E_2(\bm{p}_2)}\right.\nonumber\\[1ex]
&\hspace{11.11ex}\left.-\frac{\Lambda_1^-(\bm{p}_1)\,\gamma_0\,
[K(\bm{p},\bm{q})\,\phi(\bm{q})]\,\gamma_0\,\Lambda_2^+(\bm{p}_2)}
{P_0+E_1(\bm{p}_1)+E_2(\bm{p}_2)}\right).\label{Eq:SE}\end{align}
Multiplying the Salpeter equation (\ref{Eq:SE}) from the left by
$\Lambda_1^\pm(\bm{p}_1)$ and from the right by
$\Lambda_2^\pm(\bm{p}_2)$ and using
$\Lambda_i^\pm(\bm{p})\,\Lambda_i^\mp(\bm{p})=0$ reveals that any
of its solutions $\phi(\bm{p})$ will satisfy~the~constraints
$$\Lambda_1^+(\bm{p}_1)\,\phi(\bm{p})\,\Lambda_2^+(\bm{p}_2)=
\Lambda_1^-(\bm{p}_1)\,\phi(\bm{p})\,\Lambda_2^-(\bm{p}_2)=0\
.$$These constraints halve the number of independent components of
any solution $\phi(\bm{p})$ of the Salpeter equation
(\ref{Eq:SE}); obviously, this has important implications for
practical calculations. Generally, the Bethe--Salpeter interaction
kernel $K(\bm{p},\bm{q})$ may be decomposed into a sum of products
of Lorentz-scalar potentials $V(\bm{p},\bm{q})$ and associated
Lorentz structures. Assuming the latter to be represented by
identical Dirac matrices, generically labelled $\Gamma,$ the
action of the kernel $K(\bm{p},\bm{q})$ on the Salpeter amplitude
$\phi(\bm{q})$ requested by the Salpeter equation~(\ref{Eq:SE})~is
$$[K(\bm{p},\bm{q})\,\phi(\bm{q})]=\sum_\Gamma
V_\Gamma(\bm{p},\bm{q})\,\Gamma\,\phi(\bm{q})\,\Gamma\ .$$

\subsection{Reduced Salpeter Equation}\label{Sec:RSE}By the
charge-conjugation properties of the energy projection operators
$\Lambda_i^\pm(\bm{p}),$ the second term on the right-hand side of
the Salpeter equation (\ref{Eq:SE}) corresponds to the
negative-energy components of the Salpeter amplitude
$\phi(\bm{p})$:
$\Lambda_1^-(\bm{p}_1)\,\phi(\bm{p})\,\Lambda_2^+(\bm{p}_2)\equiv
\Lambda_1^-(\bm{p}_1)\,\phi(\bm{p})\,[\Lambda_2^-(\bm{p}_2)]^{\rm
c}.$ Assuming that it is justifiable to neglect the contribution
of this latter term relative~to~that of the first term on the
right-hand side of Eq.~(\ref{Eq:SE}) leads to the {\em reduced
Salpeter equation\/}~\cite{RSE,Lucha92C}\begin{equation}
\left[P_0-E_1(\bm{p}_1)-E_2(\bm{p}_2)\right]\phi(\bm{p})=
\int\frac{{\rm d}^3q}{(2\pi)^3}\,\Lambda_1^+(\bm{p}_1)\,\gamma_0\,
[K(\bm{p},\bm{q})\,\phi(\bm{q})]\,\gamma_0\,\Lambda_2^-(\bm{p}_2)\
,\label{Eq:RSE}\end{equation}forming, unlike Eq.~(\ref{Eq:SE}), an
{\em explicit eigenvalue problem\/} for the bound-state masses
$P_0=M.$ Formally, this reduction of the (full) Salpeter equation
(\ref{Eq:SE}) to the reduced Salpeter equation (\ref{Eq:RSE}) may
be effected by imposing as further constraint
$\Lambda_1^-(\bm{p}_1)\,\phi(\bm{p})=0$ or
$\phi(\bm{p})\,\Lambda_2^+(\bm{p}_2)=0;$ as trivial consequence of
this, all solutions $\phi(\bm{p})$ of the {\em reduced\/} Salpeter
equation (\ref{Eq:RSE}) involve only the {\em positive-energy\/}
components $\phi(\bm{p})=
\Lambda_1^+(\bm{p}_1)\,\phi(\bm{p})\,\Lambda_2^-(\bm{p}_2)\equiv
\Lambda_1^+(\bm{p}_1)\,\phi(\bm{p})\,[\Lambda_2^+(\bm{p}_2)]^{\rm
c}.$ Physically, this simplification supposes that
$P_0-E_1(\bm{p}_1)-E_2(\bm{p}_2)\ll
P_0+E_1(\bm{p}_1)+E_2(\bm{p}_2)$~or
$[P_0+E_1(\bm{p}_1)+E_2(\bm{p}_2)]^{-1}\ll
[P_0-E_1(\bm{p}_1)-E_2(\bm{p}_2)]^{-1}$ holds at the level of
expectation~values; this assumption may be justified for
semirelativistic and weakly bound heavy constituents. In the
center-of-momentum frame of the bound state, $\bm{P}=\bm{0}$
clearly implies $\bm{p}=\bm{p}_1=-\bm{p}_2.$

\subsection{Pseudoscalar Bound States}\label{Sec:PSBS}To start
with, let us focus our interest to those fermion--antifermion
bound states which are represented by a Salpeter amplitude with
the least number of {\em independent\/} components: the
pseudoscalar states. These are bound states composed of some
fermion and its antiparticle, thus characterized by $m_1=m_2=m$
and well-defined behaviour under charge conjugation; the
quantum-number assignment common to all pseudoscalar states is
$J=0$ for their total spin, $P=-1$ for their parity, and $C=+1$
for their charge-conjugation parity: $J^{PC}=0^{-+}.$ Such states
are realized in nature in form of, e.g., the pion as
quark--antiquark bound~state. Henceforth, all indices $i=1,2$
distinguishing the bound-state constituents can be dropped, since
now, for instance,
$E_1(\bm{p})=E_2(\bm{p})=E(\bm{p})=E(p)\equiv\sqrt{p^2+m^2},$~with
$p\equiv|\bm{p}|\equiv\sqrt{\bm{p}^2}.$

The most general expansion of the Salpeter amplitude
$\phi(\bm{p})$ over a complete set of Dirac matrices would, of
course, introduce 16 Salpeter components. However, as a
consequence of the peculiar projector structure of the Salpeter
equation (\ref{Eq:SE}), manifesting in the constraints mentioned
in Sec.~\ref{Sec:SE}, precisely eight of these components are
independent. Specifically, for describing pseudoscalar states just
two of the latter, labelled $\varphi_1(\bm{p})$ and
$\varphi_2(\bm{p}),$ are~relevant:
$$\phi(\bm{p})=\left[\varphi_1(\bm{p})\,\frac{H(\bm{p})}{E(\bm{p})}
+\varphi_2(\bm{p})\right]\gamma_5\ .$$In fact, the above
decomposition of $\phi(\bm{p})$ applies to {\em all\/} bound
states of a spin-$\frac{1}{2}$ fermion and a spin-$\frac{1}{2}$
antifermion for which the quantum number $s$ of the sum of the two
fermion~spins is zero, i.e., to all spin-singlet states carrying
$s=0,$ irrespective of the relative orbital angular momentum
$\ell$ of the constituents. Pseudoscalar bound states are just the
special case $\ell=0.$ The one additional constraint truncating
the full to the reduced Salpeter equation enforces the equality of
the Salpeter components $\varphi_1(\bm{p})$ and
$\varphi_2(\bm{p})$:
$\varphi_1(\bm{p})=\varphi_2(\bm{p})\equiv\varphi(\bm{p}).$ Thus,
all spin-singlet, notably all pseudoscalar, solutions of a reduced
Salpeter equation~(\ref{Eq:RSE})~will~read
$$\phi(\bm{p})=\varphi(\bm{p})\,\frac{H(\bm{p})+E(\bm{p})}{E(\bm{p})}
\,\gamma_5\equiv2\,\varphi(\bm{p})\,\Lambda^+(\bm{p})\, \gamma_5\
.$$

\subsection{Radial Eigenvalue Equations}\label{Sec:REE}In order to
follow the more or less standard route of nonrelativistic
reduction, let us assume that the Bethe--Salpeter interaction
kernel $K(\bm{p},\bm{q})$ is of {\em convolution type}, i.e., is
of the form $K(\bm{p},\bm{q})=K(\bm{p}-\bm{q}),$ which entails for
the potentials $V_\Gamma(\bm{p},\bm{q})=V_\Gamma(\bm{p}-\bm{q}),$
and that $K(\bm{p},\bm{q})$ respects spherical symmetry, i.e.,
$K(\bm{p},\bm{q})=K((\bm{p}-\bm{q})^2),$ and thus
$V_\Gamma(\bm{p},\bm{q})=V_\Gamma((\bm{p}-\bm{q})^2);$ trivially,
this means that each momentum-space potential
$V_\Gamma(\bm{p},\bm{q})$ is the Fourier transform of a
spherically symmetric configuration-space potential $V_\Gamma(r)$
depending just on the radial coordinate $r\equiv|\bm{x}|.$ In this
case, any dependence on the angular variables, encoded either in
spherical harmonics or vector spherical harmonics, can be split
off; this factorization effects the reductions of both the
Salpeter equation (\ref{Eq:SE}) \cite{Lagae92a,Olsson95} and its
reduced counterpart (\ref{Eq:RSE}) \cite{Olsson96} to equivalent
systems of coupled equations for the radial factors of all
independent Salpeter components. For a particular Lorentz
structure of the Bethe--Salpeter kernel (such~that the index
$\Gamma$ identifying the Lorentz structure may be suppressed), the
interactions experienced by bound-state constituents enter the
radial eigenvalue equations in form of Fourier--Bessel transforms
$V_L(p,q)$ ($L=0,1,2,\dots$) of the radial static
configuration-space potential $V(r),$ defined in terms of
spherical Bessel functions of the first kind \cite{Abramowitz}
$j_n(z),$ $n=0,\pm 1,\pm2,\dots$:$$V_L(p,q)\equiv
8\pi\int\limits_0^\infty{\rm d}r\,r^2\,j_L(p\,r)\,j_L(q\,r)\,V(r)\
,\qquad p\equiv|\bm{p}|\ ,\qquad q\equiv|\bm{q}|\ ,\qquad
L=0,1,2,\dots\ .$$

In the following, we intend to exemplify our technique of
constructing exact solutions to the reduced Salpeter equation by
considering only a few of those Lorentz structures
$\Gamma\otimes\Gamma$ of the interaction kernel $K(\bm{p},\bm{q})$
that are commonly used in phenomenological descriptions of
hadrons, notably, of mesons, as bound states of quarks confined by
the~strong interactions.\footnote{Since the first encounter
\cite{Olsson95,Archvadze95} of unstable solutions of Salpeter
equations with confining interactions, the stability of solutions
of different relativistic bound-state equations has been an issue
of concern \cite{Parramore95,Parramore96,Babutsidze98,
Babutsidze99,Uzzo99,Kopaleishvili01,Babutsidze03}. For confining
interactions of harmonic-oscillator form in configuration space
(in which case all bound-state integral equations reduce to
easier-to-handle differential equations in momentum space), the
conditions for stability of bound-state solutions within the
instantaneous Bethe--Salpeter framework (in the sense of their
spectrum being real, discrete, and bounded from below) have been
analyzed for those Lorentz structures of the Bethe--Salpeter
interaction kernel which are most frequently employed for the
description~of hadrons as bound states of quarks: A rigorous
analytic proof of the stability of the bound-state energy spectra
entailed has been constructed for the reduced Salpeter equation
\cite{Lucha07:StabOSS-QCD@Work07,Lucha07:HORSE,
Lucha07:SSIBSECI-Hadron07} and a particular generalization of it
\cite{Lucha07:SSIBSECI-Hadron07,Lucha07:RFPHIBSE}, formulated by
taking into account the full propagators of the bound-state
constituents \cite{Lucha05:IBSEWEP,Lucha06:GIBSEEQP-C7,
Lucha05:EQPIBSE}, obtained as the solutions of the
Dyson--Schwinger equations for the corresponding two-point Green's
functions of the bound-state constituents, and this problem has
also been discussed for the (full) Salpeter equation
\cite{Lucha08:C8,Lucha10:QCD@Work10,Lucha10:C9}.}

\subsubsection{(Full) Salpeter equation}The Salpeter equation for
bound states with spin-sum quantum number $s=0$ is equivalent to a
system of two equations for the two independent radial Salpeter
components $\varphi_1(p)$ and $\varphi_2(p).$ For pure
time-component Lorentz-vector kernels
($\Gamma\otimes\Gamma=\gamma^0\otimes\gamma^0$), this~system~reads
\begin{align*}&2\,E(p)\,\varphi_2(p)+\int\limits_0^\infty\frac{{\rm
d}q\,q^2}{(2\pi)^2}\,V_0(p,q)\,\varphi_2(q)=M\,\varphi_1(p)\ ,\\
&2\,E(p)\,\varphi_1(p)+\int\limits_0^\infty\frac{{\rm
d}q\,q^2}{(2\pi)^2}\left[\frac{m^2}{E(p)\,E(q)}\,V_0(p,q)
+\frac{p\,q}{E(p)\,E(q)}\,V_1(p,q)\right]\varphi_1(q)=M\,\varphi_2(p)\
.\end{align*}

\subsubsection{Reduced Salpeter equation}\label{Sec:REE-RSE}The
(compared to the full Salpeter equation) simple and unique
energy-projector structure of the reduced Salpeter equation
(\ref{Eq:RSE}) guarantees that each spin-singlet bound-state
solution involves just one independent radial Salpeter component
$\varphi(p),$ since $\varphi_1(p)=\varphi_2(p)\equiv\varphi(p).$
The system of coupled radial equations thus collapses to a single
radial eigenvalue equation which reads, for a kernel of
time-component Lorentz-vector Dirac structure
$\Gamma\otimes\Gamma=\gamma^0\otimes\gamma^0,$
$$2\,E(p)\,\varphi(p)+\frac{1}{2}\int\limits_0^\infty\frac{{\rm
d}q\,q^2}{(2\pi)^2}\left[\left(1+\frac{m^2}{E(p)\,E(q)}\right)V_0(p,q)
+\frac{p\,q}{E(p)\,E(q)}\,V_1(p,q)\right]\varphi(q)=M\,\varphi(p)\
,$$ for a kernel of Lorentz-vector Dirac structure
$\Gamma\otimes\Gamma=\gamma_\mu\otimes\gamma^\mu,$ describing all
the interactions by an effective vector-boson exchange between the
two fermionic bound-state constituents,
$$2\,E(p)\,\varphi(p)+\int\limits_0^\infty\frac{{\rm
d}q\,q^2}{(2\pi)^2}\left(2-\frac{m^2}{E(p)\,E(q)}\right)V_0(p,q)\,
\varphi(q)=M\,\varphi(p)\ ,$$ or, for the {\em linear combination}
$2\,\Gamma\otimes\Gamma=\gamma_\mu\otimes\gamma^\mu
+\gamma_5\otimes\gamma_5-1\otimes1$ of the kernel's
Dirac~structure,
\begin{equation}2\,E(p)\,\varphi(p)+\int\limits_0^\infty\frac{{\rm
d}q\,q^2}{(2\pi)^2}\,V_0(p,q)\,\varphi(q)=M\,\varphi(p)\
.\label{Eq:RSE-L-nzm}\end{equation}

\subsection{\boldmath Special Case: Massless Bound-State
Constituents ($m=0$)}\label{Sec:MLC}In the limit of vanishing
masses of the two bound-state constituents, i.e., in the
case~$m=0,$ the bound-state equations of Sec.~\ref{Sec:REE}
simplify, of course, still further: the Salpeter equation with
time-component Lorentz-vector kernel,
$\Gamma\otimes\Gamma=\gamma^0\otimes\gamma^0,$ reduces to the set
of equations
\begin{align}&2\,p\,\varphi_2(p)+\int\limits_0^\infty\frac{{\rm
d}q\,q^2}{(2\pi)^2}\,V_0(p,q)\,\varphi_2(q)=M\,\varphi_1(p)\
,\nonumber\\&2\,p\,\varphi_1(p)+\int\limits_0^\infty\frac{{\rm
d}q\,q^2}{(2\pi)^2}\,V_1(p,q)\,\varphi_1(q)=M\,\varphi_2(p)\
,\label{Eq:FSE-T-m=0}\end{align}whereas the single radial
eigenvalue equation emerging from the reduced Salpeter equation
reads, for interactions with time-component Lorentz-vector Dirac
structure $\Gamma\otimes\Gamma=\gamma^0\otimes\gamma^0,$
\begin{equation}2\,p\,\varphi(p)+\frac{1}{2}\int\limits_0^\infty\frac{{\rm
d}q\,q^2}{(2\pi)^2}\,[V_0(p,q)+V_1(p,q)]\,\varphi(q)=M\,\varphi(p)\
,\label{Eq:RSE-T-m=0}\end{equation}for interactions with
Lorentz-vector Dirac structure
$\Gamma\otimes\Gamma=\gamma_\mu\otimes\gamma^\mu,$
\begin{equation}2\,p\,\varphi(p)+2\int\limits_0^\infty\frac{{\rm
d}q\,q^2}{(2\pi)^2}\,V_0(p,q)\,\varphi(q)=M\,\varphi(p)\
,\label{Eq:RSE-V-m=0}\end{equation}or, for kernels of the
particularly favourable Dirac structure $2\,\Gamma\otimes\Gamma=
\gamma_\mu\otimes\gamma^\mu+\gamma_5\otimes\gamma_5-1\otimes1,$
\begin{equation}2\,p\,\varphi(p)+\int\limits_0^\infty\frac{{\rm
d}q\,q^2}{(2\pi)^2}\,V_0(p,q)\,\varphi(q)=M\,\varphi(p)\
.\label{Eq:RSE-L-m=0}\end{equation}Although we focus to reduced
Salpeter equations, we need Eq.~(\ref{Eq:FSE-T-m=0}) at an
intermediate~stage.

\section{\boldmath Configuration-Space Potentials $V(r)$ by
Inversion}\label{Sec:CSP}All the radial bound-state eigenvalue
equations recalled in Secs.~\ref{Sec:REE} and \ref{Sec:MLC},
emerging from the reduced Salpeter equation (\ref{Eq:RSE}) under
our reasonable assumption of spherical symmetry, are for
$M\ne2\,E(p)$ homogeneous linear Fredholm integral equations of
the second kind but still simple enough that, for sufficiently
sophisticated choices of the Salpeter solutions $\varphi(p),$
their underlying configuration-space potentials $V(r)$ can be
extracted by analytical means. For notational convenience, we
define the Fourier--Bessel transforms to configuration space
\begin{align*}\varphi_L(r)&\equiv{\rm i}^L\,\sqrt\frac{2}{\pi}
\int\limits_0^\infty{\rm d}p\,p^2\,j_L(p\,r)\,\varphi(p)\ ,\qquad
L=0,1\ ,\\T_L(r)&\equiv{\rm i}^L\,\sqrt\frac{2}{\pi}
\int\limits_0^\infty{\rm d}p\,p^2\,j_L(p\,r)\,E(p)\,\varphi(p)\
,\qquad L=0,1\ ,\end{align*}of both radial Salpeter amplitude
$\varphi(p)$ and free-energy part $E(p)\,\varphi(p)$ in
momentum~space. In the following, we illustrate the idea of
analytical extraction of configuration-space~radial potentials
$V(r)$ by examining explicit examples: For an appropriate ansatz
for the Salpeter solution $\varphi(p),$ by application of the
Fourier--Bessel transformation to the momentum-space bound-state
equation considered we would like to get its {\em
configuration-space representation\/} in terms of $\varphi_L(r)$
and $T_L(r);$ from the latter formulation, we should be able to
read off $V(r).$ Since in Eqs.~(\ref{Eq:RSE-L-nzm}) and
(\ref{Eq:RSE-T-m=0})--(\ref{Eq:RSE-L-m=0}) the potential may
absorb any mass $M\ne0,$ we assume $M=0.$ Of course, the simple
procedure sketched above can only be followed if the interaction
term in the bound-state equation under study contains merely a
single Fourier--Bessel transform $V_L(p,q)$ of the radial
potential $V(r),$ i.e., involves a unique value of $L.$ If, on the
other hand, both $V_0(p,q)$ and $V_1(p,q)$ enter in the
interaction term, as happens, e.g., for any interaction of
time-component Lorentz-vector structure, a different line of
reasoning has to be devised. For good reasons, we first invert
equations for massless~bound-state~constituents
(Sec.~\ref{Subsec:BSC-m=0}). Then we turn to the more delicate
case of bound-state constituents of finite mass
(Sec.~\ref{Subsec:BSC-nzm}).

\subsection{\boldmath Massless Bound-State Constituents:
$m=0$}\label{Subsec:BSC-m=0}In order to be able to deal
simultaneously with interaction kernels of Lorentz-vector nature
and of the (simplifying) linear combination yielding
Eq.~(\ref{Eq:RSE-L-nzm}), we introduce a parameter~$\eta$~by
$$\eta=\left\{\begin{array}{l}2\qquad\mbox{for
$\Gamma\otimes\Gamma=\gamma_\mu\otimes\gamma^\mu$}\ ,\\[1ex]
1\qquad\mbox{for $\Gamma\otimes\Gamma=\frac{1}{2}\left(
\gamma_\mu\otimes\gamma^\mu+\gamma_5\otimes\gamma_5-1\otimes1
\right)$}\ .\end{array}\right.$$By this definition, the reduced
Salpeter equations (\ref{Eq:RSE-V-m=0}) and (\ref{Eq:RSE-L-m=0})
may be subsumed in the~form
\begin{equation}2\,p\,\varphi(p)+\eta\int\limits_0^\infty\frac{{\rm
d}q\,q^2}{(2\pi)^2}\,V_0(p,q)\,\varphi(q)=M\,\varphi(p)\
.\label{Eq:RSE-TL-m=0}\end{equation}For this equation of motion,
its $L=0$ Fourier--Bessel transform is straightforwardly found:
$$2\,T_0(r)+\eta\,V(r)\,\varphi_0(r)=M\,\varphi_0(r)\ .$$Thus, the
configuration-space potential associated to mass eigenvalue $M=0$
is found~to~be
\begin{equation}V(r)=-\frac{2\,T_0(r)}{\eta\,\varphi_0(r)}\
.\label{Eq:CSP}\end{equation}

\newpage

In the case of interaction kernels of the time-component
Lorentz-vector Dirac structure, i.e.,
$\Gamma\otimes\Gamma=\gamma^0\otimes\gamma^0,$ the reduced
Salpeter equation involves, even for zero-mass bound-state
constituents, Fourier--Bessel transforms $V_L(p,q)$ of $V(r)$ for
more than one $L,$ namely, both $V_0(p,q)$ {\em and\/} $V_1(p,q).$
From such type of bound-state equation the potential $V(r)$ {\em
cannot\/} be recovered by applying to Eq.~(\ref{Eq:RSE-T-m=0}) a
Fourier--Bessel transformation of a particular value of $L.$ In
order to overcome this adverse observation, we recall that, as
consequence of the {\em equality\/}
$\varphi_1(p)=\varphi_2(p)\equiv\varphi(p)$ of the two independent
radial components of Salpeter amplitudes for spin-singlet bound
states enforced by the reduced-Salpeter constraint discussed in
Sec.~\ref{Sec:RSE}, for a definite Lorentz structure
$\Gamma\otimes\Gamma$ of the interaction kernel the radial
eigenvalue equation resulting from the {\em reduced\/} Salpeter
equation (\ref{Eq:RSE}) can be found by adding, for
$\varphi_1(p)=\varphi_2(p),$ the two equations that constitute the
set of (originally coupled) radial eigenvalue equations related to
the corresponding Salpeter equation (\ref{Eq:SE}); see, for
instance,~Footnote 2 of Ref.~\cite{Lucha07:HORSE}. Bearing these
findings in mind, we seek, for tentative solutions $\varphi(p)$ of
the reduced Salpeter equation (\ref{Eq:RSE-T-m=0}) for pure
time-component Lorentz-vector interaction, the responsible
potential $V(r),$ with the help of the {\em full\/}-Salpeter
``precursor'' (\ref{Eq:FSE-T-m=0}) of Eq.~(\ref{Eq:RSE-T-m=0}),
via a two-step procedure:
\begin{enumerate}\item After equating the two independent
components $\varphi_1(p)=\varphi_2(p)\equiv\varphi(p),$ we
represent the decoupled relations arising from the set of
equations (\ref{Eq:FSE-T-m=0}) in configuration space by
application of the appropriate Fourier--Bessel transformation to
each of the relations:
\begin{align*}2\,T_0(r)+V_0(r)\,\varphi_0(r)=M\,\varphi_0(r)\ ,\\[1ex]
2\,T_1(r)+V_1(r)\,\varphi_1(r)=M\,\varphi_1(r)\ .\end{align*}At
this stage we must take into account, by an index $L=0,1,$ the
possibility that the potentials $V_L(r)$ derived from each of
these relations by analogy to Eq.~(\ref{Eq:CSP})~can~differ:
\begin{equation}V_L(r)\equiv-\frac{2\,T_L(r)}{\varphi_L(r)}\
,\qquad L=0,1\ .\label{Eq:AP}\end{equation}\item We {\em assume\/}
that the unique configuration-space potential $V(r)$ we are
seeking may be expressed as a {\em linear combination\/}
$V(r)=c_0\,V_0(r)+c_1\,V_1(r)$ of the auxiliary functions $V_0(r)$
and $V_1(r),$ with yet to be determined (of course, constant)
coefficients $c_0$ and $c_1.$ We attempt to find these
coefficients by inserting our ansatz for $V(r)$ into the slightly
more complex and intricate reduced Salpeter equation
(\ref{Eq:RSE-T-m=0}) for $M=0;$ if we manage to deduce thereby a
solution for $c_0$ and $c_1,$ our quest for the
potential~$V(r)$~is~completed.\end{enumerate}

\subsubsection{Non-Confining Interaction Potentials}\label{Sec:NCP}
\begin{figure}[h]\begin{center}
\psfig{figure=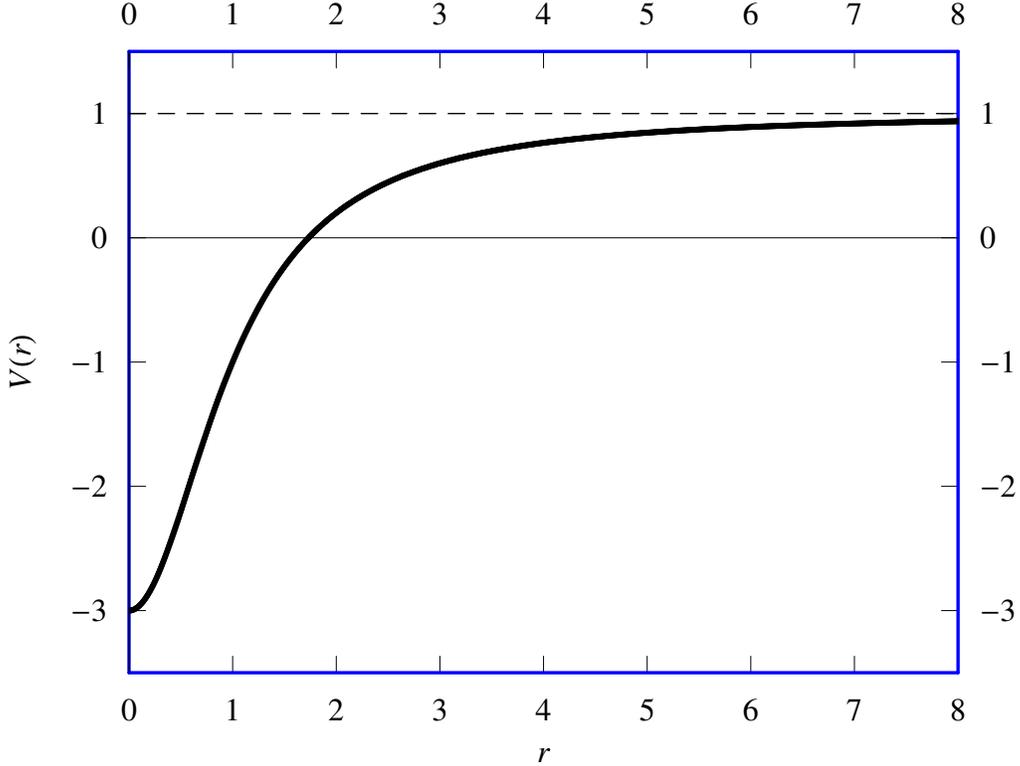,scale=1}
\caption{Configuration-space {\em non-confining\/} potential
$V(r)=1-4/(r^2+1),$ determined by inversion of the {\em reduced\/}
Salpeter equation (\ref{Eq:RSE-TL-m=0}) with kernels of {\em
Lorentz-vector\/} nature ($\eta=2$) when assuming an exponential
form $\varphi(p)\propto\exp(-p)$ of its momentum-space~solution
$\varphi(p).$ Starting at $V(0)=-3,$ this potential $V(r)$ behaves
for $r\to\infty$ like $V(r)\to1$
(dashed~line).}\label{Fig:LV-BJK-Pot} \end{center}\end{figure}

Presumably, the first guess that comes to one's mind for the
momentum-space bound-state amplitude $\varphi(p)$ is the
exponential, involving a parameter $\lambda$ with dimension of
inverse~mass:\begin{equation}
\varphi(p)=2\,\lambda^{3/2}\exp(-\lambda\,p)\ ,\qquad\lambda>0\
,\qquad\|\varphi\|^2\equiv\int\limits_0^\infty{\rm
d}p\,p^2\,|\varphi(p)|^2=1\ .\label{Eq:SA-E}\end{equation}For this
choice, the $L=0,1$ Fourier--Bessel transforms of $\varphi(p)$ and
$m=0$ kinetic term~read\begin{alignat*}{2}
&\varphi_0(r)=\sqrt{\frac{2}{\pi}}\,
\frac{4\,\lambda^{5/2}}{(r^2+\lambda^2)^2}\ ,\qquad
&\varphi_1(r)=\sqrt{\frac{2}{\pi}}\,\frac{4\,{\rm
i}\,\lambda^{3/2}\,r}{(r^2+\lambda^2)^2}\ ,\\[1ex]
&T_0(r)=\sqrt{\frac{2}{\pi}}\,4\,\lambda^{3/2}\,
\frac{3\,\lambda^2-r^2}{(r^2+\lambda^2)^3}\ ,\qquad
&T_1(r)=\sqrt{\frac{2}{\pi}}\,\frac{16\,{\rm
i}\,\lambda^{5/2}\,r}{(r^2+\lambda^2)^3}\ .\end{alignat*}If the
Lorentz structure of the interaction kernel is either a pure
vector ($\Gamma\otimes\Gamma=\gamma_\mu\otimes\gamma^\mu$) or a
linear combination of vector, pseudoscalar, and scalar
($2\,\Gamma\otimes\Gamma=\gamma_\mu\otimes\gamma^\mu
+\gamma_5\otimes\gamma_5-1\otimes1$) as summarized by the reduced
Salpeter equation (\ref{Eq:RSE-TL-m=0}), our findings
(\ref{Eq:CSP}) immediately entail, for a Salpeter amplitude
$\varphi(p)$ of the {\em exponential\/} form (\ref{Eq:SA-E}), in
configuration space the~potential$$V(r)=\frac{2}{\eta\,\lambda}
\left(1-\frac{4\,\lambda^2}{r^2+\lambda^2}\right),\qquad
V(0)=-\frac{6}{\eta\,\lambda}\ ,\qquad
V(r)\xrightarrow[r\to\infty]{}\frac{2}{\eta\,\lambda}\ .$$In the
limit $r\to\infty,$ this potential $V(r)$ approaches a finite
value. It is thus a representative of the class of non-confining
interactions, with dependence on the radial variable $r$ (in~units
of $\lambda,$ i.e., for $\lambda=1$) for Lorentz-vector
Bethe--Salpeter kernels ($\eta=2$) as depicted in
Fig.~\ref{Fig:LV-BJK-Pot}.

\noindent For the reduced Salpeter equation (\ref{Eq:RSE-T-m=0})
with time-component Lorentz-vector Dirac structure
$\Gamma\otimes\Gamma=\gamma^0\otimes\gamma^0$ and an exponential
ansatz (\ref{Eq:SA-E}) for $\varphi(p),$ the auxiliary functions
(\ref{Eq:AP})~become$$V_0(r)=\frac{2}{\lambda}
\left(1-\frac{4\,\lambda^2}{r^2+\lambda^2}\right),\qquad
V_1(r)=-\frac{8\,\lambda}{r^2+\lambda^2}\ .$$Inserting these
potentials into Eq.~(\ref{Eq:RSE-T-m=0}) fixes the coefficients
$c_{0,1}$ to $c_0=c_1=\frac{1}{2},$ which entails a
configuration-space potential $V(r)$ with behaviour clearly
similar to that shown~in Fig.~\ref{Fig:LV-BJK-Pot}:
$$V(r)=\frac{1}{\lambda}
\left(1-\frac{8\,\lambda^2}{r^2+\lambda^2}\right),\qquad
V(0)=-\frac{7}{\lambda}\ ,\qquad
V(r)\xrightarrow[r\to\infty]{}\frac{1}{\lambda}\ .$$

\subsubsection{Confining Interaction Potentials}
\begin{figure}[h]\begin{center}
\psfig{figure=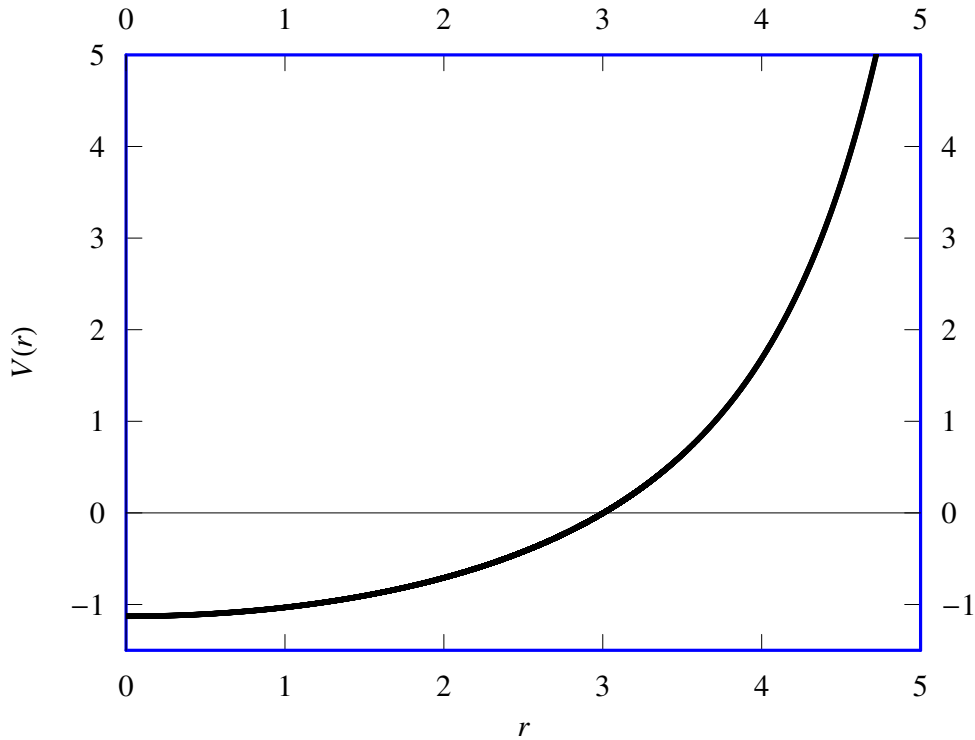,scale=.95844}
\caption{Configuration-space {\em confining\/} potential
$V(r)=\left(\frac{r}{2}-\frac{1}{r}\right)\mbox{erf\/i}\!
\left(\frac{r}{2}\right)-\,\exp(r^2/4)/\sqrt{\pi},$ found by
inversion of the {\em reduced\/} Salpeter equation
(\ref{Eq:RSE-TL-m=0}) with {\em Lorentz-vector\/} kernels
($\eta=2$) if trying a Gaussian form $\varphi(p)\propto\exp(-p^2)$
for its momentum-space solution $\varphi(p).$ Starting at
$V(0)=-2/\sqrt{\pi}\approx-1.12837\dots,$ this potential $V(r)$
behaves like $V(r)\to\infty$ for $r\to\infty.$}
\label{Fig:LV-BJK-ConfPot}\end{center}\end{figure}

In order to present
also an example for a confining potential, we next consider an
amplitude of normalized {\em Gaussian\/} form, using a parameter
$\lambda$ with dimension of inverse mass squared:\begin{equation}
\varphi(p)=2\,{\left(\frac{8\,\lambda^3}{\pi}\right)\!}^{1/4}
\exp(-\lambda\,p^2)\ ,\qquad\lambda>0\
,\qquad\|\varphi\|^2\equiv\int\limits_0^\infty{\rm
d}p\,p^2\,|\varphi(p)|^2=1\ .\label{Eq:SA-G}\end{equation}
Trivially, the $L=0$ Fourier--Bessel transform of this function
$\varphi(p)$ is also of Gaussian~form:
$$\varphi_0(r)={\left(\frac{2}{\pi\,\lambda^3}\right)\!}^{1/4}
\exp\!\left(-\frac{r^2}{4\,\lambda}\right).$$The
configuration-space potential entailed by
Eq.~(\ref{Eq:RSE-TL-m=0}) involves the imaginary error function
$\mbox{erf\/i}(z)$ defined in terms of the error function
$\mbox{erf}(z)$ \cite{Abramowitz} by $\mbox{erf\/i}(z)\equiv-{\rm
i}\,\mbox{erf}({\rm i}\,z);$ for large $r,$ it rises like
$16\,\lambda^{3/2}\exp(r^2/4\,\lambda)/\eta\,\sqrt{\pi}\,r^4$ and
thus realizes confinement, as
illustrated~by~Fig.~\ref{Fig:LV-BJK-ConfPot}:
\begin{align*}V(r)&=\frac{1}{\eta\,\lambda}
\left[\left(r-\frac{2\,\lambda}{r}\right)
\mbox{erf\/i}\!\left(\frac{r}{2\,\sqrt{\lambda}}\right)
-2\,\sqrt{\frac{\lambda}{\pi}}
\exp\!\left(\frac{r^2}{4\,\lambda}\right)\right],\qquad
V(0)=-\frac{4}{\eta\,\sqrt{\pi\,\lambda}}\ .\end{align*}

\noindent A time-component Lorentz-vector Dirac structure can be
dealt with along the same lines as in Subsec.~\ref{Sec:NCP};
however, the resulting expressions are lengthy and not really
enlightening.

\subsection{\boldmath Bound-State Constituents with Non-Zero Mass:
$m>0$}\label{Subsec:BSC-nzm}The extension of the particular
inversion technique framed here to the case of {\em
nonvanishing\/} masses $m$ of the bound-state constituents clearly
requires somewhat more careful choices of momentum-space
amplitudes $\varphi(p).$ One's first attempt might employ the
rational function\begin{equation}\varphi(p)
=\sqrt{\frac{2}{\pi}}\,\frac{4\,\mu^{5/2}}{(p^2+\mu^2)^2}\
,\qquad\mu>0\ ,\qquad\|\varphi\|^2=\int\limits_0^\infty{\rm
d}p\,p^2\,|\varphi(p)|^2=1\ ,\label{Eq:SA-R}\end{equation}with a
mass parameter $\mu.$ Its $L=0$ Fourier--Bessel transform is, for
$\mu\le m,$ an
exponential:$$\varphi_0(r)=2\,\mu^{3/2}\exp(-\mu\,r)\
,\qquad0<\mu\le m\ .$$For illustrative purposes, we apply this
Salpeter amplitude to invert the particularly simple reduced
Salpeter equation (\ref{Eq:RSE-L-nzm}) with the linear combination
$\Gamma\otimes\Gamma=\frac{1}{2}\,(\gamma_\mu\otimes\gamma^\mu
+\gamma_5\otimes\gamma_5-1\otimes1)$ of Lorentz structures. In
order to evaluate the relevant Fourier--Bessel transform of
Eq.~(\ref{Eq:RSE-L-nzm}),\begin{equation}
2\,T_0(r)+V(r)\,\varphi_0(r)=M\,\varphi_0(r)\qquad\Longrightarrow\qquad
V(r)=-\frac{2\,T_0(r)}{\varphi_0(r)}\qquad\mbox{for $M=0$}\
,\label{Eq:IP}\end{equation}we need, for $m\ne0,$ the $L=0$
Fourier--Bessel transform $T_0(r)$ of the kinetic part
$E(p)\,\varphi(p).$

\subsubsection{\boldmath Case $0<\mu=m$}
\begin{figure}[t]\begin{center}
\psfig{figure=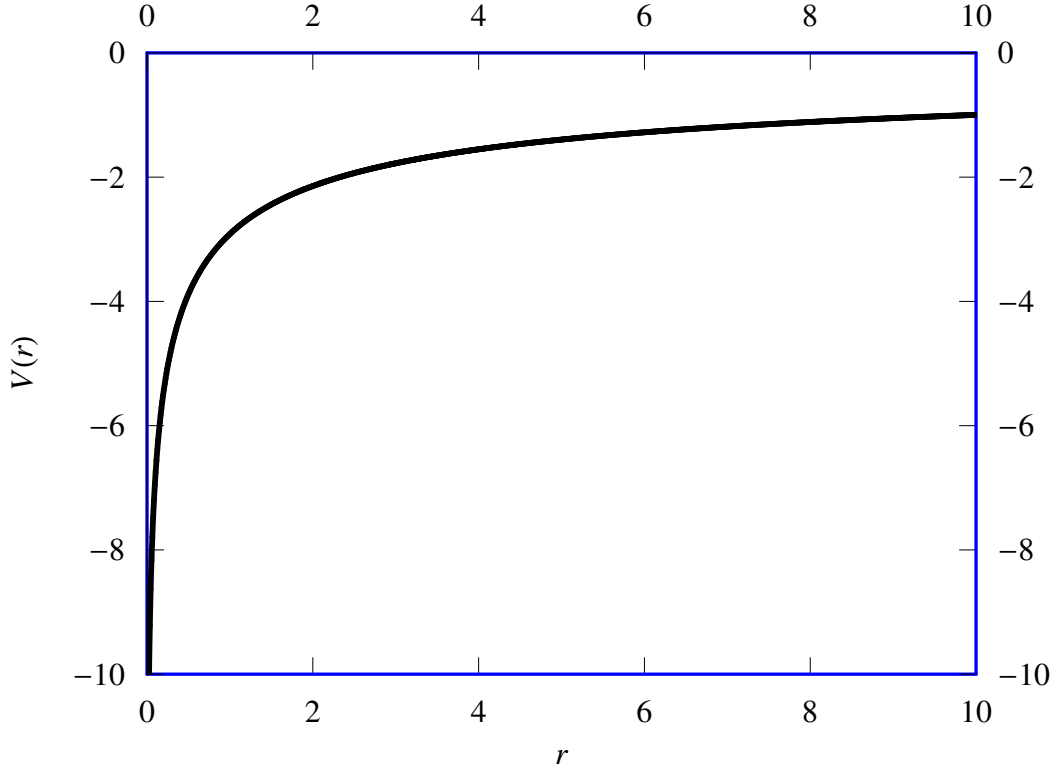,scale=1}
\caption{Configuration-space {\em non-confining\/} potential
$V(r)=-(8/\pi)\,K_0(r)\exp(r),$ found by studying the {\em
reduced\/} Salpeter equation (\ref{Eq:RSE-L-nzm}) for bound-state
constituents of {\em nonvanishing\/} mass $m$ with a kernel of
Lorentz structure
$\Gamma\otimes\Gamma=\frac{1}{2}\,(\gamma_\mu\otimes\gamma^\mu
+\gamma_5\otimes\gamma_5-1\otimes1)$ when relying on a rational
function $\varphi(p)\propto(p^2+1)^{-2}$ as ansatz for the
momentum-space solution~$\varphi(p).$ This $V(r)$ vanishes in the
limit $r\to\infty$ but exhibits a logarithmic singularity at
the~origin~$r=0.$}
\label{Fig:BJK-Bessel-Pot}\end{center}\end{figure}

For the rational Salpeter
amplitude (\ref{Eq:SA-R}), the $L=0$ Fourier--Bessel transform of
the kinetic part is easily pinned down if the parameter $\mu$ is
chosen to be equal to the constituents' mass $m;$ the outcome for
$T_0(r)$ is basically the modified Bessel function $K_\nu(z)$
\cite{Abramowitz}~of~order~$\nu=0$:
$$T_0(r)=\frac{8\,m^{5/2}}{\pi}\,K_0(m\,r)\ .$$Hence, the instant
reply of our inversion procedure (\ref{Eq:IP}) is the
configuration-space potential
$$V(r)=-\frac{8\,m}{\pi}\,K_0(m\,r)\exp(m\,r)\ ,\qquad
V(r)\xrightarrow[r\to0]{}\frac{8\,m}{\pi}\ln(m\,r)\ ,\qquad
V(r)\xrightarrow[r\to\infty]{}0\ .$$The monotonic increase of this
potential $V(r),$ from its logarithmic singularity at the~origin
$r=0$ to its asymptotic value 0 for $r\to\infty,$ is depicted for
constituent mass $m=1$ in Fig.~\ref{Fig:BJK-Bessel-Pot}. The
singularity is milder than Coulombic and hence compatible with
Herbst's~findings \cite{Herbst}.

\subsubsection{\boldmath Case $0<\mu<m$}Allowing the parameter $\mu$
in our ansatz (\ref{Eq:SA-R}) for the Salpeter amplitude
$\varphi(p)$ to be less than the constituents' mass $m,$ i.e., if
$\mu<m,$ performing a contour integration and adopting the residue
theorem enables us to cast the resulting configuration-space
potential into~the~form$$V(r)=-2\left\{\sqrt{m^2-\mu^2}
+\frac{\mu}{\sqrt{m^2-\mu^2}}\,\frac{1}{r}-\frac{4\,\mu}{\pi\,r}
\int\limits_m^\infty{\rm d}\rho\,\rho\exp[-r\,(\rho-\mu)]\,
\frac{\sqrt{\rho^2-m^2}}{(\rho^2-\mu^2)^2}\right\}.$$For
$r\to\infty,$ the $r$-dependent terms of the potential vanish and
$V(r)$ approaches a constant:
$$V(r)\xrightarrow[r\to\infty]{}-2\,\sqrt{m^2-\mu^2}\ .$$Examining
the $r$-dependent portion of this potential $V(r)$ for $r\to0$ by
L'H\^opital's rule, we encounter a logarithmically divergent
integral. Hence, since for $\mu\uparrow m$ the asymptotic~value of
$V(r)$ reduces to zero, the behaviour of $V(r)$ resembles that one
found for the case $\mu=m.$

\section{Summary, Conclusions, and Outlook}In this study, we showed
how to establish, by analytical means, exact relationships between
solutions of Bethe--Salpeter equations and the underlying ---
instantaneous --- interactions: for elaborate assumptions about
the nature of both interactions and resulting bound states
culminating in a manageable structure of the equations governing
such bound states, this is accomplished by postulating particular
solutions and reading off the interaction potentials. Among
others, having these rigorous solutions at one's disposal
obviously~provides a useful or even decisive test when solving
Bethe--Salpeter equations numerically by conversion~into
equivalent {\em matrix eigenvalue problems\/} (as, e.g., in
Refs.~\cite{Lagae92a,Olsson95,Olsson96,Lagae92b,Bonn94,
Lucha00:m0,Lucha00:C4,Lucha00:n0,Lucha01:IAS}) or when attempting
to construct {\em approximate models\/} for Bethe--Salpeter
solutions, as proposed in Refs.~\cite{HL11,HL12}.

Clearly, the three-dimensional Fourier transform of any function
which depends only on a radial coordinate is just the $L=0$
Fourier--Bessel transform of this function. Accordingly, due to
the simplicity of the bound states inspected some but not all of
our reduced~Salpeter equations, more precisely,
Eqs.~(\ref{Eq:RSE-L-nzm}), (\ref{Eq:RSE-V-m=0}), and
(\ref{Eq:RSE-L-m=0}), are, in fact, equivalent to so-called
spinless Salpeter equations with, where necessary, appropriately
adjusted overall coupling strength (for reviews on this latter
bound-state equation see, e.g.,
Refs.~\cite{Lucha92,Lucha94:Como,Lucha04:TWR}). Earlier attempts
to construct exact solutions of spinless Salpeter equations may be
found in, e.g.,
Refs.~\cite{Brau99,Lucha99,RHO05,Chargui09,Kowalski}.

In order to provide a kind of ``proof of feasibility'' of the
inversion technique constructed here, this formalism has been
elaborated only for the simplest conceivable problem, that is, the
one posed by the {\em reduced\/} Salpeter equation. There exist,
however, exceptional cases for which the above findings apply
directly, without changes, also to the {\em full\/} Salpeter
equation: As recalled in Sec.~\ref{Sec:PSBS}, bound states with a
vanishing sum of the spins of their constituents, such as
pseudoscalar states, are represented by only two independent
Salpeter components, the minimal number of independent components
for solutions of the {\em full\/} Salpeter equation.
Correspondingly, for these states the full Salpeter equation
becomes equivalent to a system of merely two, in general coupled,
equations. For a Lorentz structure of the Bethe--Salpeter
interaction kernel of, for example, the form
$2\,\Gamma\otimes\Gamma=\gamma_\mu\otimes\gamma^\mu
+\gamma_5\otimes\gamma_5-1\otimes1,$ one of the latter equations
does not contain any interactions and is therefore of purely
algebraic nature \cite{Lucha07:HORSE}. For vanishing bound-state
mass, the two equations decouple, and the inversion~problem for
the full Salpeter equation thus becomes identical to that for the
reduced Salpeter~equation.

The intention behind this study was to carry out an analysis of
purely academic~nature. Nevertheless, one may ask the legitimate
question: to which physical bound states observed in nature do the
above considerations apply? Section \ref{Sec:RSE} confines the
validity of the {\em reduced\/} Salpeter equation to
semirelativistic, weakly bound, heavy constituents; this
precludes, for instance, the pion but not necessarily pseudoscalar
mesons composed of heavy quarks, such as $\eta_c$ and $\eta_b,$
nor the $h_c$ and $h_b$. The range of application to be expected
for any~{\em full\/}~Salpeter equation is, of course, much wider.
However, a thorough study of the latter equation as well as the
extraction of a realistic potential are definitely beyond the
scope of the~present~work.

\end{document}